# New Insight into Collimated Plasmon Beam: Nondiffracting versus Linear Focusing

L. Li, T. Li,* S. M. Wang, and S. N. Zhu

National Laboratory of Solid State Microstructures,

School of Physics, College of Engineering and Applied Sciences, Nanjing University,

Nanjing 210093, China,

**We worked out a new group of collimated plasmon beams by the means of in-plane diffraction with symmetric phase modulation. As the phase type changes from 1.8 to 1.0, the beam undergoes an interesting evolution from focusing to a straight line. Upon this, an intuitive diagram was proposed to elucidate the beam nature and answer the question whether they are nondiffracting or linear focusing. Based on this diagram, we further achieved a highly designable scheme to modulate the beam intensity (e.g., "lossless" plasmon). Our finding holds remarkable generality and flexibility in beam engineering and would inspire more intriguing photonic designs.**

*Corresponding authors, E-mails: taoli@nju.edu.cn

URL: http://dsl.nju.edu.cn/litao

Surface plasmon polariton (SPP) is a combined excitation of electron and photon, which provides a possible solution to integrate the light in sub-wavelength scale and suggests many charming applications due to its enhanced local field and light-matter interaction [1]. Among various SPP manipulations, nondiffracting beam is both of fundamental interest and useful functionality (e.g., guiding surface wave). As the counterpart of optical Airy beam in free space [2,3], the exciting realization of Airy plasmon provides a good solution to achieve nondiffracting surface wave without any nonlinearity [4-8]. In fact, beam engineering has recently attracted increasing interests, and was further boosted to arbitrary convex trajectories [9,10] and non-paraxial region [11-13]. Another aspect of highly interested is the beam focusing that reflects the capability of people to concentrate optical field, which has already received extensive studies in plasmonic regime and was even developed to SPP demultiplexers [14-17]. In some sense, nondiffracting beam can be regarded as a particular kind of focusing whose focal spot is preserved in beam propagation, and is more worth to engineer elaborately. However, the well demonstrated SPP Airy beam would not be always convenient in on-chip integrations for its bending trajectory. To generate a straight SPP beam with strong field confinement and controllable intensity will be of great importance.

In principle, straight nondiffracting plasmon beam can be formed by interference of two intersecting plane waves with linear wave-front phase (just like the generation of Bessel beam in free space [18]), which was indeed realized by Lin *et al.* very recently [19]. However, this linear phase is quite different from the 1.5-power phase type of nondiffracting Airy beam. By carefully investigating them together with the previous focusing case [17], we would find it should attribute to a distinct feature of the Airy beam - asymmetric phase modulation as 1.5-power phase in *x*<0 while exponential decay in *x*>0 [20], which also explains the beam asymmetry of self-bending. Thereafter, a straightforward question arises that how about using a symmetric 1.5-power phase to generate a straight nondiffracting SPP beam? If not, what is the contribution of different phase and what kind of beam will be constructed?

In this letter, we utilize the non-perfectly-matched (NPM) Bragg diffraction method [7] to establish arbitrary SPP beams along *z*-axis with symmetric phase modulation as type from 1.8 to 1.0. Self-collimated SPP beams are well characterized by leakage radiation microscopy (LRM), which is a well developed technique using oil-immersed objective to extract the SPPs of large *k*-vectors [21,22]. In some cases, these straight SPP beams appear to be less-dispersive till collapse at certain distances. On the other hand, they also look like the elongated focal spots. Thereafter, we raise a paradox of the beams: nondiffracting or focusing to a line? An intuitive diagram is proposed to illustrate the physical insight into the formation of such kind of beams with respect to different phase modulations. Moreover, an interesting "lossless" beam with particular intensity evolution is realized. Based on the proposed diagram, the modulation on beam intensity is further generalized in a highly designable scheme, which holds remarkable generality and flexibility in steering plasmon beams. The new insight and full understanding of the beams are highlighted and possible applications are discussed.

The strategy to modulate the SPP beam phase by NPM diffractions of non-periodic nano-cave array has been illustrated in detail [7]. Shortly, when an SPP beam passes through a nano-array, it will be diffracted to various directions by local units for constructive interference, resulting in an extra $2\pi$ phase change between every neighboring row. By careful design, a well diffracted SPP beam with desired lateral phase modulation is accessible. Figure 1(a) schematically shows the SPP diffraction by non-periodic array, where different optimum diffractions correspond to different units with different angles $\theta$ for constructive interference. According to the phase distribution in $x$-axis, $\phi_m(x)=\phi_0+k_{spp}x-2m\pi$ ($\phi_0$ is an reference phase of the incident wave and $m$ is the sequence number of the lattice point), we are able to achieve any required phase type in a general form as $\psi(x)=-cx^n$.

First, we would start from a symmetric $n=1.5$ phase modulation by extending the nano-array much longer than that of previous Airy beam, which covers a wider range of the local lattice in $x$-direction ($a_x$) from 826 nm to 477 nm. The detailed parameters are depicted in Fig. 1(c), which were calculated by solving $\phi_m(x)=\psi(x)$ (with a proper coefficient $c=0.6$). The retrieved symmetric phase distribution along $x$-axis shown as symbols in Fig. 1(d) is well consistent with the designed one (solid curve). Having these data, we fabricated the sample by focus ion beam (Strata FIB 201, FEI Company) milling on a silver film with ~60 nm thickness on a quartz substrate. Figure 1(b) shows scanning electron microscopy (SEM) image of the top-view of the array sample, where the unit is a rectangular nano-hole with size of 240×120 nm$^2$, depth of 20 nm, period ($z$ direction) of $P_z$= 610 nm and local lattice in Fig. 1(c). A white arrow in the figure indicates the position of the local lattice matches Bragg condition ($a_x$~610 nm), which is defined as the symmetric point ($x$=0) and the length of array in ±$x$ directions are almost the same (both ~9.7 $\mu$m).

Experiment was carried out with an illumination of He-Ne laser ($\lambda_0$=632.8 nm), and SPP wave was launched by a grating with period of 610 nm in corresponding to the SPP wavelength. Figure 2(a) shows the LRM recorded bidirectional SPP beams diffracted by the array, revealing apparent straight collimation of the center lobes. To give more detailed information, beam intensities of cross-sections within the selected region [in Fig. 2(a)] are plotted in Fig. 2(b), where a center non-dispersive peak is well demonstrated with the FWHM kept ~1.5 $\mu$m within ~30 $\mu$m distance till it vanishes. Interestingly, the peak intensity experiences a slight increase to an abrupt drop as shown in the inset image. Theoretical calculation was subsequently performed on an SPP beaming with a well designed $n$=1.5 phase design with an SPP attenuation length of ~15 $\mu$m [7], as the result shown in Fig. 3(a). Both the beam trajectory and center lobe profile (inset image) well reproduce experimental results (stronger oscillations may attribute to cleaner interference). Since the designed phase is symmetric, it is ready to accept that constructive interference always occurs in the center line. However, this interesting beam intensity profile with almost "lossless" within 30 $\mu$m and a following abrupt drop would likely imply the underlying physics and provide insights to get a full understanding of this collimated SPP beam.

As has been illustrated in NPM Bragg diffraction [7], different local lattice will lead to an optimum diffraction angle $\theta$ [see Fig. 1(c)]. This angle can be derived from the differential of the phase function [23], as

$$\sin\theta = -\frac{1}{k}\frac{\partial \psi(x)}{\partial x} = \frac{cn}{k}x^{n-1} \qquad (1)$$

by adopting $\psi(x) = -cx^n$. Thereafter, we can easily get the relation of a beam region in $z$-axis with respect to the contribution from the sources region (diffraction units) in $x$-axis as

$$z = x\cot\theta = \sqrt{(\frac{k}{cn}x^{2-n})^2 - x^2} \ . \qquad (2)$$

Calculating Eq. (2) with $n=1.5$ and $c=0.6$, we obtained a diagram of the correspondence curve of beam in $z$-axis and source in $x$-axis, as shown in Fig. 3(b). An apparent increasing slope of the curve indicates more sources contribute as the beam propagates, which explains the beam tends to be enhanced with an increasing trend. As for the almost flat intensity profile in experiment, it would attribute to SPP loss that suppresses this enhancement. On the other hand, since the designed array is finite (-9.7<$x$<9.7 $\mu$m), there is no source out of this range [indicated by dashed lines in Fig. 3(b)] to contribute to the beam, and thus leads to an abrupt drop in beam intensity at $z\sim34$ $\mu$m.

For a straightforward generalization, we explored other cases of the phase type changing from $n=1.8$ to $1.0$. Figures 4(a-d) show the results of samples with fixed $c=0.6$ and $n=1.8, 1.6, 1.4$ and $1.2$, respectively. A clear evolution from a strong focusing to a nearly nondiffracting beaming is demonstrated. Notably, the case of $n=1.0$ with $c=1.5$ is particularly shown in Fig. 4(e) (*c* is changed for the image visualized better), where a nondiffracting beam is clearly observed with the preserved beam shapes shown in Fig. 4(f). Corresponding to these designs, we calculated their source-to-beam diagrams in order to get an in-depth recognition of these beams, as the results shown in Fig. 4(g). Except the linear case of *n*=1.0, others all have an increase tendency in their *x-z* curves, and the lager *n*, the larger increment. Same to the previous *n*=1.5 case, there is a cutoff in the main lobe due to the finite array region, as indicated by a dashed line of *x*=9.7 $\mu$m. It does explain the experimental results. Howbeit, for the case of *n*=1.8, we will find that no matter how large of the source is provided the beam range is limited, and most sources only contribute to a narrow region around *z*~10 $\mu$m. In this regard, it appears likely to be a focusing process. Whereas, does it mean that the smaller *n* cases truly correspond to infinite beaming as long as the sources are infinite?

Revisiting Eq. (2), we will find it can be renormalized for some particular cases. For example, when *n*=1 the it degenerates to a line with slope of $\sqrt{(\frac{k}{cn})^2-1}$; when *n*=1.5 it becomes a circle equation with radius of $r=\frac{1}{2}(\frac{k}{1.5c})^2$ and centre of (*r*, 0); when *n*=2 it is still a circle with *r*=*k*/2*c* and centre of (0, 0) [24]. It is clear that the *x-z* curve for *n*=1.5 in Fig. 3(b) is just a part of the circle arc, which implies the main lobe of beam will not go farther than the radius (*z*=*r*=65.4 $\mu$m) even if the source is infinite. As for other cases, the calculated correspondence *x-z* curves reveal that their curvatures are determined by the coefficient *c* and phase factor *n*. The same point is that they all only cover a finite region except *n*=1.0 [24]. In fact, the *n*=1.0 phase type truly corresponds to the case of two intercrossing SPP plane wave with the result of nondiffracting Cosine-Gauss beam [19]. Again in Fig. 4(g), the straight *x-z* line of *n*=1.0 case does indicate an equal contribution of sources to the beam everywhere, which is also a nature of nondiffracting beam. However, the experimental result definitely shows us a decay beam, which is rightly due to the plasmonic loss as well as in Ref. [19]. Now, it is well recognized that beams of the phase type *n*≠1 are not non-diffraction since they are all limited within a certain range. However, the small *n* cases have revealed the nearly nondiffracting characteristic and weaker side lobes if they are carefully designed. In this regard, these self-collimated beams are able to work as the nondiffracting one to some extent. It does provide a robust method to generate a straight SPP beam with narrow beam width and almost non-diffraction, which holds more generality for further manipulations.

Furthermore, a highly designable scheme is proposed to retrieve a proper irregular phase to achieve a collimated beam with any desired intensity profile. As we know, SPP wave always suffers from the large propagation loss even for the nondiffracting cosine-gauss beam [19]. Our design has manifested the capability to compensate the loss by phase design in a certain range, as demonstrated in Fig. 2 [or for the case of *n*=1.4 in Fig. 3(c)]. Indeed, the phase can be retrieved by strictly deducing a proper source contribution to compensate the propagation loss and thus achieve an exact intensity preserved SPP beam. According to above discussions, we introduce a source-to-beam density as $\rho \propto \dfrac{dx}{dz}$, where $z$ is related with $x$ via an optimum angle $\theta$ [see Eq. (2)]. Assuming an SPP with attenuation length of $l$ and a fixed intensity of unit source, the beam intensity in $z$-axis contributed by a unit source can be expressed as

$$I_i \propto \exp(-\dfrac{r}{l}) = \exp(-\dfrac{x}{l\sin\theta}). \tag{3}$$

Thereafter, we can artificially tune the source-to-beam density to get a constant beam intensity as $I(z) = \rho I_i = c$. Then, we have an equation of

$$\dfrac{dx}{dz}\exp(-\dfrac{x}{l\sin\theta}) = c. \tag{4}$$

Solving Eqs. (1) (2) and (4), we will obtain the phase distribution along $x$-axis for a loss compensated SPP beam. Here, we theoretically suppose the SPP wave have an attenuation length of $l=20$ $\mu$m. A particular phase distribution can be numerically retrieved [see Fig. 5(a)] to realize a straight SPP beam with intensity preserved as the result shown in Fig. 5 (b). The inset image depicts a flat intensity profile (although a little oscillation) of center lobe revealing a good "lossless" feature within a range of $z<50$ $\mu$m (red curve). It is also apparently compared with the intact nondiffracting cosine-Gauss beam with an exponential decay (white curve), showing a well achieved intensity preserved SPP beam!

More interestingly, this scheme can be further generalized by introducing an arbitrary function F($z$), as $I(z) = \rho I_u = c\mathrm{F}(z)$, to modify the beam intensity profile almost at will. An easily conceivable example is an exponential increasing beam with the function of $\mathrm{F}(z) = \exp(\dfrac{z-z_0}{l'})$, where $l'$ is a coefficient and $z_0$ is a start point. Figure 5(c) depicts the designed phase with respect to $l'$=20 $\mu$m and $z_0$=10 $\mu$m, exhibiting a more curved phase compared with the "lossless" one. Calculation results is shown in Fig. 5(d) revealing a good exponential increase in the region of $z$<40 $\mu$m, which well shows the high controllability of our design. It is reasonably believed that our design is capable of achieving more complicated intensity profiles with an arbitrary function of F($z$) as long as large enough sources are provided, which would possibly imply particular applications (e.g., in optical trapping [26]). Notably, this designable scheme is not only valid for the lossy SPP but also for other systems as the item of exp(-$r$/$l$) in Eq. (3) is replaced by a more general function (e.g., it may be used to suppress a light beam in a gain medium).

In addition to these novelties addressed above, these collimated SPP beams retain other unique properties, e.g., the self-healing property, which was also experimentally observed [24]. It should be noted that owing to the in-plane NPM diffraction process all advantages of previous Airy beam [7] and focusing [17] are inherited in this newly realized SPP beams, such as independence to SPP launcher, broadband property, and so on.

In summary, we have successfully achieved a group of self-collimated SPP beam by symmetric phase modulation from the phase type of $n$=1.8 to 1.0 using the NPM technique. A paradox between a nondiffracting versus linear focusing was proposed and insensitively studied. Utilizing an intuitive source-to-beam diagram, we successfully explained the characteristics of these beams. It is concluded that although some of these beams have nondiffracting appearance, they are not the real ones except the particular case of *n*=1.0. As a result of constructive diffractions, these SPP beams can be regarded as a kind of the focusing with respect to a line but not a point. This newly discovered source-to-beam relationship offers us a powerful and flexible method to design an intensity controllable beam. Our study gives a unique insight into the collimated SPP beam formation and be expected to inspire more intriguing phenomena and potential applications in beam engineering and nanophotonic manipulations.

This work is supported by the State Key Program for Basic Research of China (Nos. 2012CB921501, 2010CB630703, 2009CB930501 and 2011CBA00200) and the National Natural Science Foundation of China (Nos. 11174136, 10974090, 11021403 and 60990320) and PAPD of Jiangsu Higher Education Institutions..

**Figure Captions**

FIG. 1. (Color online) (a) Scheme of generating collimated SPP beam by in-pane diffractions. (b) Top-view of the FIB fabricated non-periodic nano-hole array, where an arrow is marked to indicate the symmetric center of the phase modulation. (c) Designed local lattice data ($a_x$) by solving equation of $\psi(x)=-cx^n$ for the $n$=1.5 case. (d) Retrieved phase distribution in lateral dimension of the beam (square symbol) compared with the designed phase type (solid curve).

FIG. 2. (Color online) (a) LRM recorded SPP beam propagations as diffracted by the well designed nano-array. (b) Detailed beam cross-sections within the selected region in (a), inside which an inset image depicts the beaming profile of the center main lobe intensity, revealing an increase-to-drop tendency.

FIG. 3. (Color online) (a) Theoretical calculation of SPP propagation beam with symmetric phase type $n=1.5$, where the inset image shows the intensity profile of center main lobe. (b) Source-to-beam correspondece curve revealed in $x$-$z$ diagram, in which the dashed blue line indicates the region of the finite sources.

FIG. 4. (Color online) (a-e) LRM recorded SPP beam with the diffracted phase modulation of $n=1.8$, 1.6, 1.4, 1.2, and 1.0, respectively ($c=1.5$ for $n=1.0$ and $c=0.6$ for others, and all in same scale). (f) Recorded beam shapes ($n=1.0$) at different propagation distances (10, 15, 20, 25 $\mu$m, from up to bottom). (g) Souce-to-beam correspondence curves for all cases, in which the finite source region is indicated by blue dashed line.

FIG. 5. (Color online) (a) Retrived the phase distribution along $x$-axis of an intensity preserved SPP beam, and (b) the corrsponding calculation result. The inset image plots intensity profile of center lobe (red curve) with a comparsion with a calculated cosine-Guass one (white curve). (c) Phase of an exponential increasing beam with (d) the calculation result.

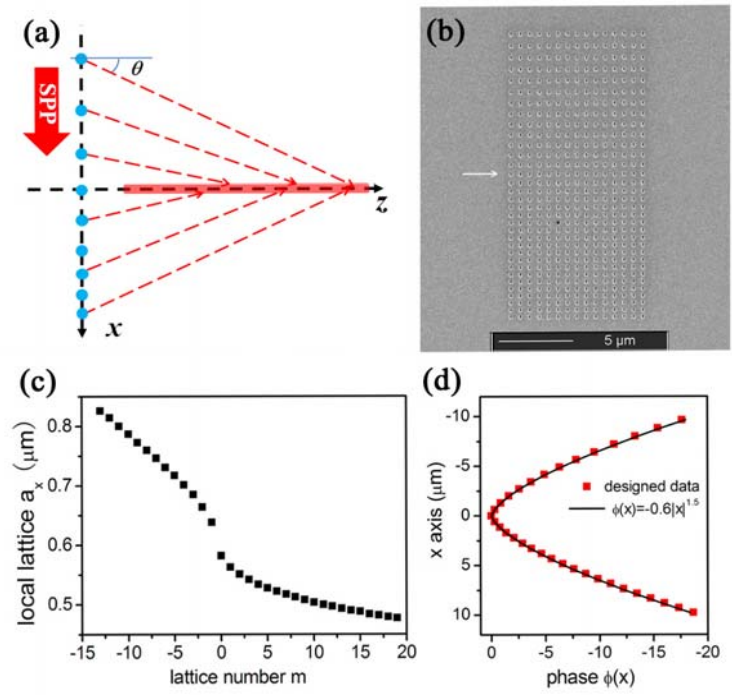

FIG. 1

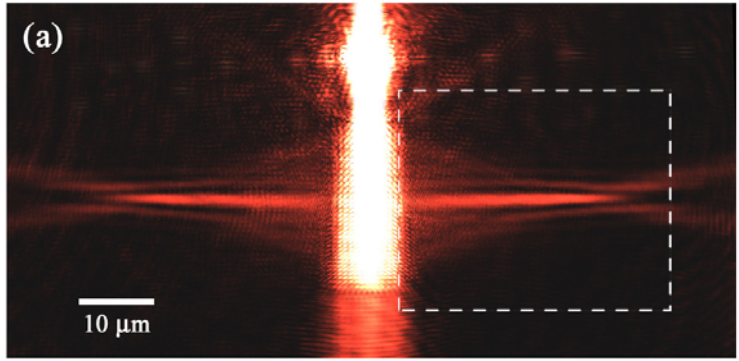

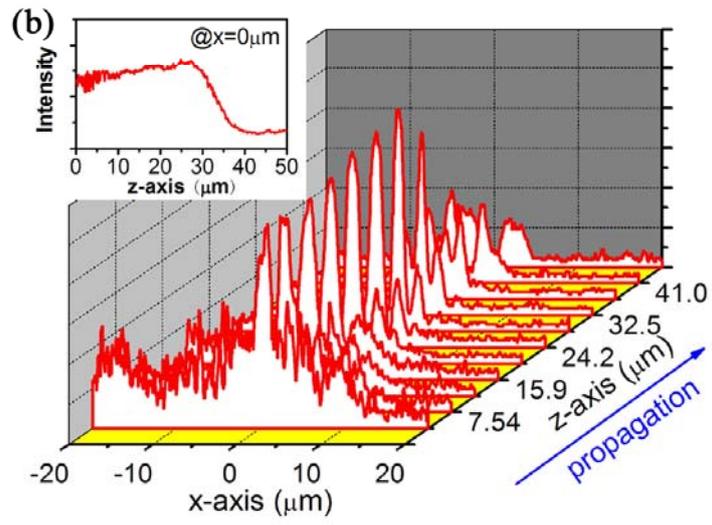

FIG. 2

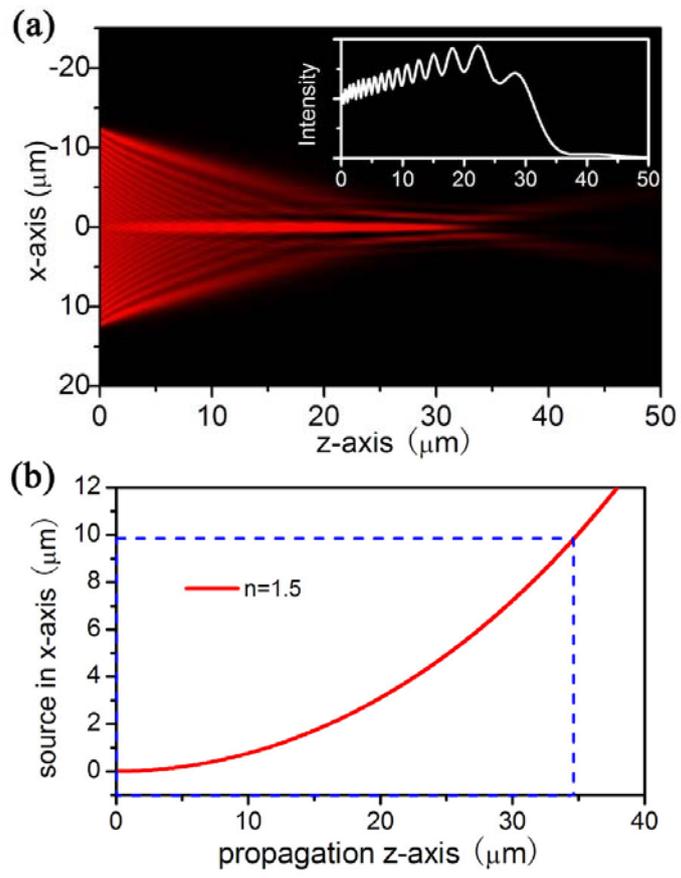

FIG. 3

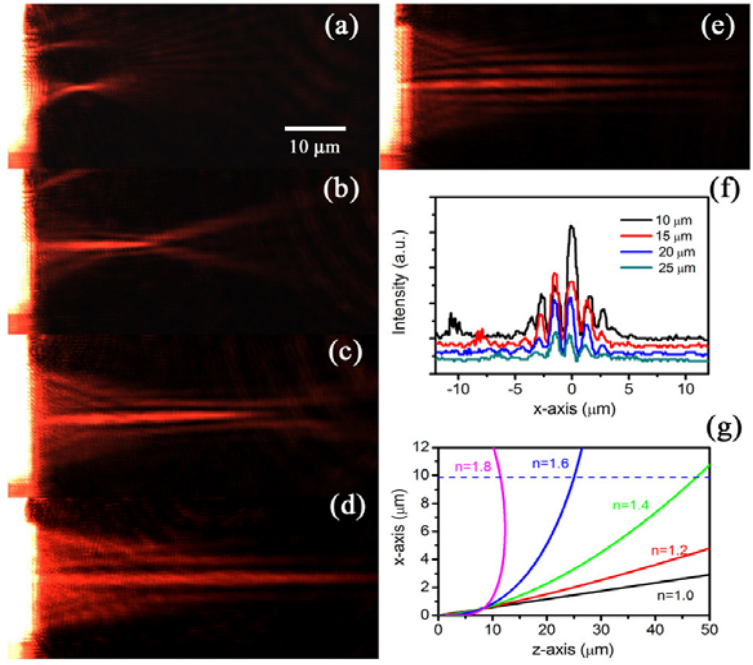

FIG. 4

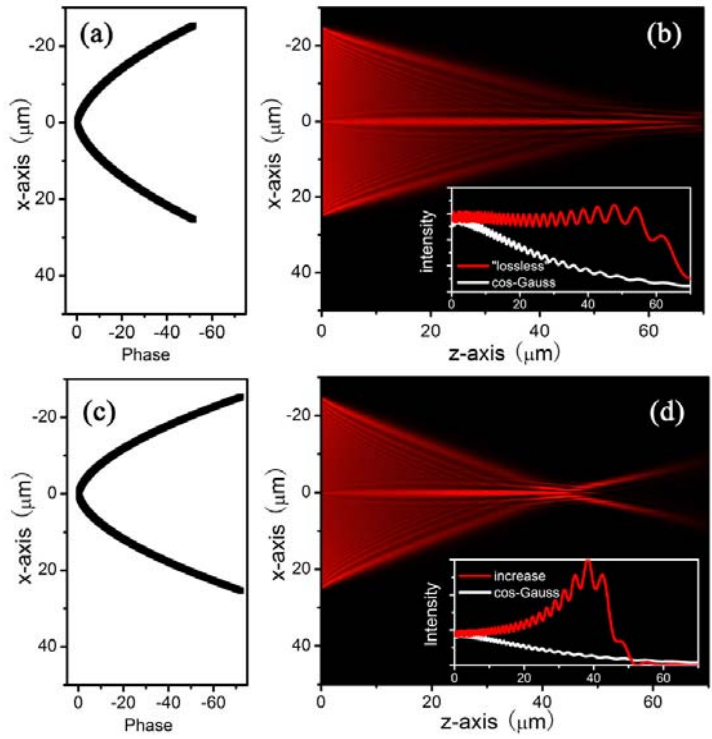

FIG. 5

# Supplementary information

**1. Diagram of source-to-beam correspondence curves**

We propose a diagram to illustrate the contribution of the nano-array diffraction units (source) on the collimated center SPP beam (along the *z*-axis). According to Eq. (2), we are able to carefully analyze different line types of this source-to-beam correspondence (*x-z*) curve. Firstly, we would like to formalize this equation for some particular cases as:

(i) $z = \pm x\sqrt{(\frac{k}{cn})^2 - 1}$, $(n=1)$;  (S1)

(ii) $z^2 + [x - \frac{1}{2}(\frac{k}{cn})^2]^2 = \frac{1}{4}(\frac{k}{cn})^4$, $(n=1.5)$;  (S2)

(iii) $z^2 + x^2 = (\frac{k}{cn})^2$, $(n=2)$.  (S3)

It can be clearly seen that for case-(i), it is a straight line with the slope of $\sqrt{(\frac{k}{cn})^2 - 1}$. As for case (ii) and (iii), they are both circles with different center locations and radii. Also, we can find the radius of circle in case-(ii) is much larger than case-(iii). For example, with the coefficient *c*=0.6 (in our experiments), $r_{(ii)} = \frac{1}{2}(\frac{k}{cn})^2 = 65.49\ \mu m$ is much larger than $r_{(iii)} = (\frac{k}{cn}) = 11.44\ \mu m$. Figure S1(a) depicts the *x-z* correspondence curves for these three cases for a vivid illustration. Such a large contrast in scale is also a reason that we haven't presented the result of *n*=2.0 in context.

As for other cases, although they cannot be analytically formalized to any well-defined line types, it is possible to calculate their *x-z* correspondence curves numerically. As has been illustrated, two factors determine the curvature: coefficient c and phase power *n*. To demonstrate these curves in a proper range, different c is selected corresponding to different *n*, as the results shown in Fig. S1(b). It is observed all types of curves turn back at large *x* region within limit range in *z*-axis, corresponding to finite diffractive beams. It should be noted, unlike the results in Fig. 4(g) in the context, the smaller *n* cases seem to have larger tangent value, because their coefficient *c* have been intentionally enlarged. A major difference of these beams with respect to the *n*=2.0 focusing is their *x-z* curves all start from the origin point, indicating these beams must extend from the zero point whatever more or less.

On the contrary, a typical focus should only have an *x-z* curve in a limited *z*-axis projection around the focal spot away from the origin point. In fact, the quadratic phase type (*n*=2) commonly used for focusing is just in a paraxial approximation, which also can be revealed in our *x-z* curve diagram as shown in Fig. S2. The black half-circle is the result of *n*=2 and *c*=0.6. For this case, if the source area covers a range of -10<*x*<10 $\mu$m, the correspondence beam may exist within a wide range as 0<*z*<8.5 $\mu$m where the circle arc covers. It is not a focus apparently. If we shrink the source area, e.g., -2<*x*<2 $\mu$m (indicated by two dashed lines), the arc projects a narrow region in *z*-axis corresponding to a good focusing, as the paraxial condition is satisfied. According to our experiments, the samples were fabricated ~20 $\mu$m long (-10<*x*<10 $\mu$m). If one wants to get a good focusing, a larger focus length should be designed by decreasing the *c* value (e.g., *c*=0.2, the red arc in Fig. S2), which also leads to a narrow range in *z*-axis projection (around *z*=25 $\mu$m).

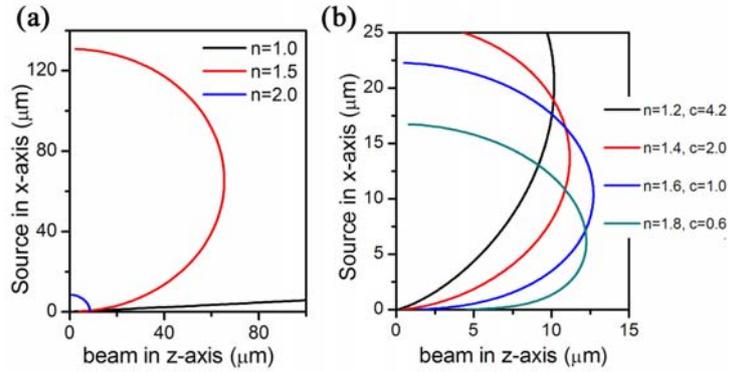

FIG. S1

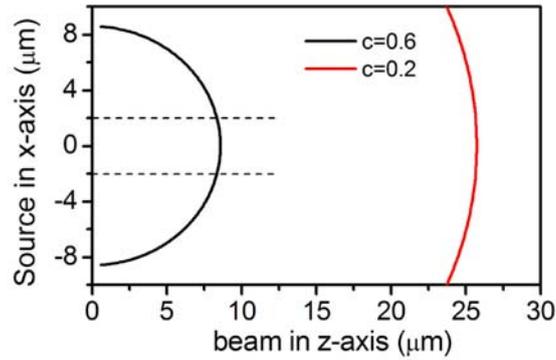

FIG. S2

2. **Self-healing property of the collimated SPP beam**

We would like to demonstrate the self-healing property of the beam with $n=1.4$. Figure S4 demonstrate the experimental results, where big holes (6~8$\lambda$ in size) were intentionally introduced in the SPP beam path to block (a) the side sub-lobes and (b) the center main lobe. It is clearly observed that both beams (left) are almost self-healed as compared with the intact ones on the other side (right). Probably, we may be confused by this self-healing property according to the explanation of the source-to-beam correspondence diagram. Since every beam part has a corresponding source part, the beam will be changed if any part of the source is blocked. In fact, the results in Fig. S3 truly provide us this information. We may find the beam main lobe in longer distance of (a) is weaker than that of (b), and vice versa in the near parts. It should be mentioned that source-to-beam diagram is an ideal consideration described by geometric optics. In reality, the NPM Bragg diffraction is based on constructive interference and local lattice would inevitably lead to some dispersive diffractions that brings a divergence angle $\Delta\theta$. This divergence results in the center beam intensity may not strictly determined by the corresponding sources in $x$-axis. The nearer to the center of the sources (in $x$-axis), the larger range of beam (in $z$-axis) will be covered due to this divergence; and at the center point ($x=0$), the source should cover the whole $z$-axis.

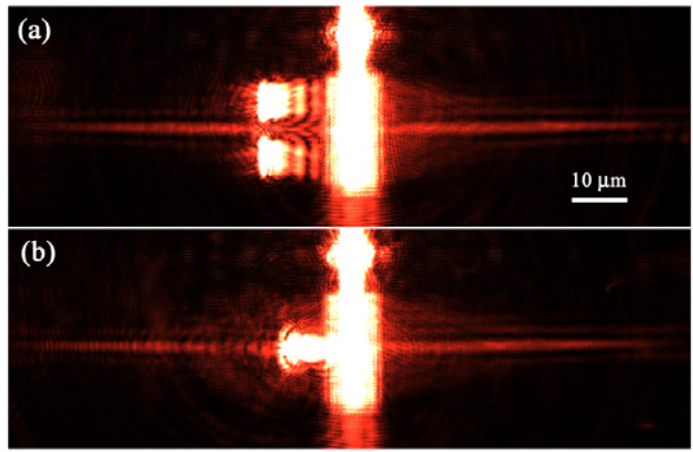

FIG. S3